\documentclass[aps,preprint,a4paper]{revtex4} 
\newcommand{\be}{\begin{equation}}
\newcommand{\ee}{\end{equation}}
\newcommand{\bea}{\begin{eqnarray}}
\newcommand{\eea}{\end{eqnarray}}
\usepackage{graphicx}
\def\Journal#1#2#3#4{{#1} {\bf #2}, #3 (#4)}

\def\BiJ{ Biophys. J.}                
\def\Bios{ Biosensors and Bioelectronics}

\def\JCP{ J. Chem. Phys.}
\def\JAP{ J. Appl. Phys.}
\def\JMB{ J. Mol. Biol.}

\def\NLE{{ Nature Lett.}}

\def\PRL{ Phys. Rev. Lett.}

\def\PRB{{ Phys. Rev.} B}

\def\RMP{ Rev. Mod. Phys.}
\def\EPJD{{ Eur. Phys. J.} D}
\def\SAB{ Sens. Act. B}

\def\eR{\textbf{\em R} }

\def\aa{\AA\,}
\begin{document}
\title{A network model to investigate structural and electrical 
properties of proteins}

\author{Eleonora Alfinito}

\affiliation{Dipartimento di Ingegneria dell'Innovazione, 
Universit\`a del  Salento\\
Via Arnesano, I-73100, Lecce, Italy,\\
Consorzio Nazionale Interuniversitario per le Scienze Fisiche
della Materia (CNISM)}
\thanks{Corresponding author e-mail: eleonora.alfinito@unile.it}

\author{Cecilia Pennetta}
\affiliation{Dipartimento di Ingegneria dell'Innovazione, Universit\`a del 
Salento\\
Via Arnesano, I-73100, Lecce, Italy,\\
Consorzio Nazionale Interuniversitario per le Scienze Fisiche
della Materia (CNISM)}

\author{Lino Reggiani}
\affiliation{Dipartimento di Ingegneria dell'Innovazione, Universit\`a 
del Salento\\
Via Arnesano, I-73100, Lecce, Italy,\\
Consorzio Nazionale Interuniversitario per le Scienze Fisiche
della Materia (CNISM)}

\begin{abstract}

One of the main trend in to date research and development is the 
miniaturization of electronic devices. 
In this  perspective, integrated nanodevices based on proteins or 
biomolecules are attracting a major interest. In fact, it has been shown that proteins like bacteriorhodopsin and azurin,
manifest electrical properties which are promising for the 
development of active components in the field of molecular electronics. 
Here we focus on two relevant kinds of proteins: The
bovine rhodopsin, prototype of GPCR protein, and the enzyme 
acetylcholinesterase (AChE), whose inhibition is one of the 
most qualified treatments of Alzheimer disease. 
Both these proteins exert their functioning starting with 
a conformational change of their native structure. Our guess is that 
such a change should 
be accompanied with a detectable variation of their electrical properties.
To investigate this conjecture, we present an impedance network model 
of proteins, able to estimate the different electrical response 
associated with the different configurations.
The model resolution of the electrical response is found able to monitor
the structure and the conformational change of the given protein. 
In this respect,
rhodopsin exhibits a better differential response than AChE. 
This result gives room to different interpretations of the degree of
conformational change and in particular supports a recent hypothesis 
on the existence of a mixed state already in the native 
configuration of the protein. 

\end{abstract}

\date{\today}

\maketitle

\section{Introduction}
Proteins are at the basis of living systems. 
In particular, they control the sensing action and the correct 
functioning of the system  by operating at a cellular level. 
In a cell, the starting process of this functional control is usually
associated with a conformational change of the protein in response to 
the capture of a specific ligand.
The conformational change activates a sequence of biological 
mechanisms, which end in the production of a final stimulus used by the system 
to organize its life. 
The possibility to detect such a conformational change through 
some modification of an electrical property of the corresponding
protein is of extreme interest from both a fundamental 
and an applied point of view.

On the former side, the nature of charge transport in 
protein \cite{Dutton} and, in general, in biological matter 
\cite{Weiss}, is a long time explored subject till now not completely 
understood \cite{Frau}. 
Most of the models are  based on a quantum mechanical tunneling 
\cite{Frau}. On this respect, some lean towards a charge transport
along the protein backbone \cite{Onuchic}, others point to a "rest and
fire" thermal mechanism  in which charge is injected only for certain values
of the dihedral angles \cite{SHE00}.

On the latter side, applications, the use of proteins, and in general
of biomolecules, for the construction of nanodevices is one of the leading
frontiers in biology, physics and technology. 
In fact, there are various concomitant peculiarities 
which point to biomolecules 
as promising material for new devices. Among them, we remember their 
extremely small sizes, their specificity, 
i.e. the ability to react only to a specific stimulus (or a very 
narrow window of similar stimuli), their conductive properties \cite{rinaldi03,jin06}, and not negligible,
their low cost \cite{Din01}.

In this context, the most intriguing problem is to understand how 
conformational changes influence the electrical properties of a 
protein. 
At present, most of the information about the different 
structures of proteins comes from X-ray or NMR 
investigations \cite{PDB}. 
The polypeptide is crystallized (X-ray analysis) 
and then a map of all its atoms is acquired. 
Starting from these representations it is possible to model the 
protein by using, either an all-atom scheme like in Molecular Dynamics 
techniques \cite{Go}, 
or an all-amino acid scheme like in Elastic 
Network models \cite{Ben,Tirion,Atilgan,Maritan}.
The former method is surely more accurate but needs for a big 
computational environment, the latter method requires more affordable
computational resources and anyway is able to catch the relevant 
features of the polypeptide.
Many authors have described the topological and statistical properties 
of this kind of network, also inducing physical information in fine 
accordance with the experimental 
tests \cite{Ben,Tirion,Atilgan,Maritan}. 
\par
The aim of this paper is to develop an irregular network model, 
which translates 
a given protein topology into the protein electrical properties. 
Accordingly, starting from the three-dimensional representation of the 
protein, the model will carry out a comparative analysis of its 
geometrical structures and predict some features of its small signal 
impedance.    
Calculations are specialized to bovine rhodopsin, the prototype
of G-protein coupled receptors (GPCR), and to the enzyme AChE, 
whose inhibition is one of the most qualified treatments of Alzheimer
disease. By means of Nyquist plots, the electrical response of the 
corresponding 
networks are then investigated for different values of frequency of the 
applied external field. 
\par
The content of the paper  is organized as follows.
Section II briefly reviews the theoretical approach.
Section III reports the results obtained for the two proteins considered
here by analyzing the topological network and the impedance spectra
for the known structures available in the literature.
Major conclusions are drawn in Sec. IV. 
Finally, the Appendix details the values of the protein equivalent 
circuit obtained from the simulations.

\section{Model}

Since the seminal paper of Tirion \cite{Tirion}, many researchers 
have investigated the possibility to model proteins by means of 
appropriate networks \cite{Atilgan,Maritan}.
This coarse grained approach is 
computationally more affordable than a molecular dynamics 
procedure \cite{Go} and, nevertheless, able to simulate the global 
behavior of a given protein \cite{Tirion,Atilgan,Maritan}.
Within this approach, a protein is mapped into a network whose 
nodes coincide with the coordinates of the  so-called 
C$_\alpha$ carbon atom pertaining to each amino acid as known from
the public data base (PDB) \cite{PDB}.
As free parameter, the model contains  the interaction (cut-off) 
radius, \eR, which fixes the maximal distance between two connected 
nodes: Only nodes with distance less or equal to \eR are connected 
with a link. 
This simple model provides a map of the protein spatial organization, 
i.e. a topological network or a graph. 
It becomes pregnant of meaning only when the links take a physical 
role. 
This was made, for example, by assigning a spring constant to each 
link and thus studying the network normal modes \cite{Tirion,Atilgan,Maritan}. 

In some recent works \cite{UPON04,WYL05,ICNF05}, by assigning to each
link an impedance, we proposed a new strategy for the investigation 
of the protein electrical response.
Furthermore, on the same ground, also the role of thermal 
fluctuations was tested \cite{UPON04,WYL05,ICNF05}.
Here we use the same strategy by assuming that the amino acids 
interact  electrically among them: The electrical charge 
transfers between neighboring residues \cite{Yang}, 
which can also change their electronic polarization \cite{Song}. 
More precisely, in this model each link between the nodes $i$ and $j$ 
corresponds to an elementary 
RC parallel impedance $Z_{i,j}$, with R an Ohmic resistor and and C
a planar homogeneous capacitor.  
By assuming that the cross-sectional area of resistor and capacitor 
is equal to the cross-section area of overlapping spheres, the 
elementary impedance becomes \cite{WYL05}
\begin{equation} 
Z_{i,j}={l_{i,j}\over {\mathcal{A}}_{i,j}}   
{1\over (\rho^{-1} + i \epsilon_{i,j}\, \epsilon_0\omega)} \label{zeta} 
\end{equation}
where ${\mathcal{A}}_{i,j}=\pi ($\eR$^2 -l_{i,j}^2/4)$, is the cross-sectional
area between the  spheres centered on the $i,j$ nodes, respectively,
$l_{i,j}$ is the distance between these centers, $\rho$ is the 
resistivity, taken to be the same for every 
amino acid with the indicative value of an insulator 
$\rho = 10^{10} \ \Omega$ m; $i$ is the imaginary unit, $i=\sqrt{-1}$, 
$\epsilon_0$ is the vacuum permittivity
and  $\omega$ is the circular frequency of the applied voltage. 
The relative dielectric constant of the couple of $i,j$ amino 
acids, $\epsilon_{i,j}$, is expressed in 
terms of the intrinsic  polarizability of the $i,j$ amino acids, 
$\alpha_{i},\alpha_{j}$  \cite{WYL05,Song} as:
\begin{equation}
\epsilon_{i,j} = 1+g[\frac{(\alpha_{i}+\alpha_{j})}{{2}}-1]
\end{equation}
where g=4.65 is a normalization constant introduced to obtain values 
of $\epsilon_{i,j} $ distributed between 1 and 80 (relative 
permittivity of vacuum and water at 20 $^{\circ}$C, respectively).
\par
By taking the first and last amino acid as injector and collector 
contacts at the given electrical potential difference,
the calculation of the total impedance of the network is obtained
by solving  the electrical circuit, i.e. we have found the potential 
on each node and the current flowing through each link for a given 
applied external voltage \cite{Rammal}.     
To this purpose we made use of the standard procedure of solution of 
linear networks based on Kirchhoff's laws. 
More precisely, in the present case of an irregular network 
with complex topology, it was particularly convenient to write 
and solve the Kirchhoff's node equations. 
\par
This one-node impedance model, henceforth also called  AA model, 
is here further implemented  by considering also the case of having
two nodes for each amino acid, henceforth also called AB model.
In fact, by identifying the amino acid with its C$_\alpha$,
in the one-node model the resulting
structure becomes analogous to the polypeptidic backbone. 
However, in this way only the backbone behavior can be reproduced while,
in a conformational change, the backbone displacement is not the 
only relevant transformation.
For example, one could be interested to study the rotations
of each amino acid around the backbone \cite{SHE00}. 
To account for these possibilities, we have to depart from  
the one-node model, and look for a more realistic picture.  
Since the distinctive mark of each amino acid is in its residue, which 
does not lye  on the polypeptidic backbone, we found 
natural to fix on each amino acid a second node. 
Accordingly,  as second node we choose the so-called
C$_\beta$ atom, i.e. the second carbon atom that attaches to the functional
group. 
The C$_\beta$ atom is present in all the amino acids with the exclusion
of  glycine. 
The impedance attributed to the new links arising from the presence
of this second node is taken of the same form as in Eq. (1).
In this way, the total number of nodes ($\nu$) mapping the protein 
is practically doubled. 
To connect the nodes we have adopted two choices, namely: The isotropic 
network and the directed network, this latter choice being able to 
better exploit directional characteristics in analogy with the directed
percolation \cite{Odor}.  
In the isotropic network,  the $\alpha$ and $\beta$ nodes 
are considered to be equivalent. 
Accordingly, each node is connected with all the others inside the 
interaction radius.
Thus, for sufficiently large \eR, each node has $\nu$-1 connections. 
In the directed network, the $\alpha$ and $\beta$
nodes are not equivalent.
Accordingly, for a given interaction radius,
each $\alpha$ node, identified by 
the serial number of the protein primary structure,
is connected to the $\beta$ node of its amino acid and to the $\alpha$
and $\beta$ nodes of amino acids with higher serial number. 
By contrast, each $\beta$ node is linked to $\beta$ 
nodes pertaining to amino acids with higher serial number. 
\par
Accordingly, for the isotropic network, the maximum value of 
the total number of links is: 
\be
N_{isotropic}^{max} = (N_{\alpha} + N_{\beta})( N_{\alpha} 
+ N_{\beta} -1)/2,
\label{Ni}
\ee
while, for the directed network it is:
\be 
N_{directed}^{max} = (N_{\alpha})( N_{\alpha} -1)/2 + 
(N_{\beta})( N_{\beta} -1)/2 +\sum_{\alpha}\sum_{\beta \geq \alpha} 
N_{\alpha} N_{\beta}.
\label{Nd}
\ee
Here  $N_{\alpha}$ is the number of C$_{\alpha}$, coincident with the total
number of amino acids, $N$, and $N_{\beta}$ is the 
number of C$_{\beta}$.
We observe that the considered proteins have, respectively, 
 348 $C_\alpha$ atoms and 325 $C_{\beta}$ atoms (rhodopsin) and 
527 $C_\alpha$ atoms and 483 $C_{\beta}$ atoms (AChE), so 
we can assume $N_{\beta}\,\approx\,N$
and then deduce the expressions:
\bea
N_{isotropic}^{max}& \approx & 2N(2N-1)/2, \nonumber \\
N_{directed}^{max}& \approx & N(3N-1)/2.
\label{Na}
\eea
By comparing Eqs. (\ref{Ni}), (\ref{Nd}), (\ref{Na}) we notice
a global different functioning of the isotropic and directed networks.
In the former network, each node represents an independent unity
that interacts in the same way with all the other unities by elongating 
$ (N_{\alpha} + N_{\beta} -1)\,\approx\, 2N-1$ links.
In the latter network, each amino acid represents a working unit 
that  interacts with other units through only three kinds of links, 
respectively: C$_{\alpha}$ - C$_{\alpha}$,
C$_{\alpha}$ - C$_{\beta}$, and C$_{\beta}$ - C$_{\beta}$. 
The number of links drawn out by each amino acid is $\approx\, 3N-1$.  
We notice, that the isotropic network  exhibits a number of links 
that is in general larger than the directed network up to a maximum
value of about $25 \%$.
\par
In what concerns with the contacts to the external bias, in the AA 
model these are positioned on the first and last amino acid. 
In the AB model we have explored three possibilities. 
Accordingly, as injector and collector nodes we have taken: 
(i) the first C$_\alpha$ and the last C$_{\beta}$, 
AB$_{\alpha, \beta}$ model; 
(ii) the couple $\alpha-\beta$ of the
first amino acid, and the couple $\alpha-\beta$ of the last amino-acid, 
AB$_{\alpha \beta, \alpha \beta}$ model; 
(iii) the first C$_\alpha$ atom and the last C$_\alpha$ 
atom, AB$_{\alpha, \alpha}$ model.
\section{Results}
In the following, we investigate the structural and electrical 
properties of a protein through both the graph model and the impedance 
network model. 
Then, we compare the electrical responses of a protein before 
and after undergoing a conformational change  by using 
the AA and the AB models.
The proteins under test are the bovine rhodopsin in dark (native)
and light (activated), and the enzyme acetylcholinesterase (AChE) in 
the native form and complexed form with Huperzine A.   
The former is the most explored structure in the class of the GPCR 
and is widely used as a prototype for deducing the structure and 
function of all the other proteins pertaining to the same class \cite{Gether,Santosh}. 
The latter plays a fundamental role in the
process of functioning of the muscle cells. 
It destroys the neuro-transmitter acetylcholine after it has passed to 
these cells the information coming from the brain, so that
new signals can be transmitted. 
Starting from the atomic coordinates of the given protein, as 
reported in the public data base (PDB) \cite{PDB}, we first investigate 
the topological properties of the  networks as function of the 
interaction radius.
Then, the analogous investigation is carried out for the total impedance
spectrum of the corresponding impedance network. 
\subsection{The topological network}
Bovine rhodopsin and torpedo acetylcholinesterase are quite different 
proteins, both in structure and in functioning. 
The former is a seven-$\alpha$-helices transmembrane protein, acting as 
a light receptor. 
It is able to capture photons, then producing a cascade process that
starts with the activation of a G-protein and ends with the 
transmission of information to brain. 
The latter is a globular protein, made of fourteen $\beta$-sheets and sixteen
$\alpha$-helices.
It is  an enzyme that breaks the neurotransmitter
acetylcholine into acetic acid and choline, thus stopping the transmission
of signal from nerve cells to muscle cells.
\par
At present, in the PDB \cite{PDB} there are 26 entries related to bovine 
rhodopsin and 56 entries related to torpedo acetylcholinesterase. 
For rhodopsin, only two of the 26 entries describe the protein in 
light, namely:
1LN6, an NMR product, and the most recent 2I37, obtained with X-ray 
analysis. 
None of the above two entries reports the entire sequence of 
amino acids. 
For the rhodopsin in dark (the native state)
there are three complete sequences, namely:
1U19, 2G87, 2HPY, all obtained by X-ray diffraction, with different 
resolution. 
The AChE is reported in many, incomplete,
different representations, either native or complexed (the activated 
state) with different 
molecules (mainly Tacrine,Rivastigmine, Galantamine, Huperzine A/B). 
\par
The first task we address is to asses the level of resolution  the
present topological network model can reach when  discriminates among 
similar (native or activated) representations and between the 
native and activated state. 
In other words, we would check whether the network provides a 
sensitive  map of the protein structure. 
To this purpose, the analysis is performed by using the AA model.
\par
Figure 1 reports the difference in the number of  
links between couples of structures of rhodopsin, native (left) and 
activated (right), considered here as a function of the interaction radius.
The native (Rho) and activated (MetaII) configurations refer to 
engineered structures obtained from the incomplete 1LN6, 1JFP and 1F88 
(chain A) \cite{UPON04,WYL05,ICNF05}
and contain the same number of amino acids (348). 
The comparison among the three native representations of rhodopsin, 
1U19, 2G87, 2HPY and the engineered one, Rho, shows that: 
(i) 1U19 and 2G87 exhibit practically the same  number of links
and, (ii) there are consistent differences between
the native and engineered configurations. 
Furthermore, when moving from the native to the activated 
state, we found an increase of the difference in the 
number of links. 
\par
Figure 2 reports the difference in the number of 
links between couples of native (left) and  complexed (right) 
representations of  AChE as a function of the interaction radius.
Here, the model correctly predicts that two distinct 
chains of the same representation of native AChE (1EA5$_{a/b}$) keep 
practically the same  number of links independently of the 
interaction radius.
On the other hand, representations of native AChE obtained under 
different experimental conditions show significative differences in 
the number of links, with a maximum value around \eR = 25 \aa. 
Some structures of complexed AChE are compared with the native form 
2ACE: The difference in the link number can be positive or negative 
with respect to the structure considered, and the maxima 
differences are comparable with those between couples of native 
structures. 
By comparing rhodopsin with AChE, we notice that
the values of the difference in the number of links 
and their dependence upon \eR 
are very similar, either in the native or in the activated state. 
However, we remark that AChE with its 527 amino acids
(2ACE) is a protein substantially greater than
rhodopsin with its 328 amino acids.   
As a consequence, the  relative value of the difference in the 
number of links is found to be more consistent for rhodopsin 
than for AChE.  
\par
From figures 1 and 2, it is also evident that the representations of a given
protein in the same state (native or activated) look sometimes 
comparable with those of representations of the same protein in 
different states.
In other words, different experimental conditions may 
produce very different representations of the same state of the
protein. 
Therefore, if the network model is used to discriminate between the 
native and the activated (or complexed) state of the protein, then,
it is mandatory that the reference representations should be 
produced under the same experimental conditions. 
For this reason, in the following we use the couple
Rho - MetaII, for native and activated rhodopsin, and 2ACE - 1VOT-2 
(X-ray products, same experiment) for native and complexed 
(with Huperzine A) AChE. 
The 1VOT-2 structure is the amino-acid sequence 1VOT deprived of two 
amino acids, ALA536, CYS537,  which are not present in 2ACE. 
\par
To emphasize the network model ability to catch the protein topology, we
have calculated the adjacency matrix \cite{Barabasi} by representing 
the links in a x-y plane where the serial number of protein
amino acids is reported on the x and y axes. 
Here each link corresponds to a point. 
Accordingly, Fig. 3 reports the map of rhodopsin in dark, Rho: 
The helix-to-helix links are very evident, either  for \eR = 6 \aa 
(dark points) or, even better, for \eR = 12 \aa (grey boxes). 
We find that the links reproduce the closeness of H2 with H3 and H1, 
and of H2 and H4; furthermore they suggest the presence of H-bonds 
among H1,\,H2,\,H7 and H2,\,H3,\,H4 and also between
H3 and H6 and between H6 and H7 \cite{Santosh}.
Notice that the helix couples (H1-H3), (H1-H4), (H1-H5),
(H1-H6), (H2-H5), (H2-H6),(H2-H6), (H4-H7),(H5-H7) are not connected 
for these values of the interaction radius.
\par
Figure 4 reports the map of AChE.
The adjacency matrix has been calculated for  \eR = 6 \aa 
(dark points), and for  \eR = 12 \aa (grey boxes).
Here we notice an inhomogeneous distribution of links, which 
is more tangled than in rhodopsin, mainly due to the
higher complexity of the AChE protein. 
It is impossible, within a single figure, to report all the connections 
among sheets and helices. 
Thus, we emphasize only some of the them, the most evident, mainly 
reproducing the closeness between $\beta$-sheets. 

\par 
We conclude that the graph analysis of proteins provides a valuable 
sketch of the force connected regions in the protein. 
In fact, by considering short range forces and using \eR as the parameter
describing their cut-off distance, Figs. 3 and 4 identify
the interacting regions of the protein. 
The increase of \eR is equivalent to consider forces with longer range. 
Accordingly, the drawings in Figs. 3 and 4 emphasize the dependence from
the interacting radius of the connective map of the network.
\subsection{The impedance network}
By attributing to each link of the network an elementary impedance 
as given by Eq. (\ref{zeta}), we have calculated the frequency
response of the impedance network, $Z(f)$, in the frequency range 
0$\div$1100 Hz, and represented it by means of 
the Nyquist plot. 
This kind of plot is very used in signal processing
and in characterizing the electrical properties of biological
materials appropriately deposited on functionalized gold electrodes \cite{Hou}.
It combines the imaginary and real part of a transfer function, 
in our case the impedance associated with the protein, 
using the frequency as an implicit variable. 
By means of the Nyquist plot, we compare the electrical
response of the protein in its native and activated states for different
values of the interaction radius.
\par
The AA model contains one free parameter in the value of the cut-off 
radius \eR, which fixes the number of links and so the 
network topology. 
In the limit of \eR values too small (to say \eR $<$ 6 \aa), 
only the nearest neighbours are connected, and so it is not possible 
to reveal the existence of more complex structure like $\alpha$-helices 
or $\beta$-sheets. 
On the other hand, in the limit of \eR values too large (to say \eR $>$ 80
\aa), each node is connected with all the others, and so the protein appears
as a uniform structure.
A value of \eR that is relevant for our purposes should be that 
which enables the main structures of the protein to emerge clearly. 
In fact, we are interested in detecting if and how they displace in 
the protein conformational change. 
Accordingly, we look for a value of \eR best revealing the
main structures of protein but also emphasizing the differences between
the activated and the native state of the protein.
For a GPCR, a relevant value of \eR is a compromise between
the characteristic dimension of the $\alpha$-helices and the 
typical distance among $\alpha$-helices, say D. 
When rhodopsin goes from the quiescent to the active state, its 
$\alpha$-helices change their relative distance and D goes in 
D$^{\prime}$. 
In the corresponding network, when the value of \eR is between D and 
D$^{\prime}$, a huge number of links change their value, so well 
revealing the conformational change. 
We call D the "effective distance", and use it as a reference 
length value.
\par
In order to explore the different topologies associated with
the changing of \eR,  in the frame of the AA model, 
we have evaluated the network degree distribution,
i.e. the distribution of the connected nodes \cite{Barabasi}.
The results of calculations are reported in Fig. 5 for rhodopsin 
and in Fig. 6 for AChE.
Here we observe that for \eR $\le$ 9 \aa, both for rhodopsin 
and AChE, the degree distribution remains substantially peaked
around the same degree value, 
i.e., there is a single characteristic dimension 
of the network clustering. 
It corresponds to the nearest neighboring
domain (k=6 for AChE, k=7 for Rho). 
The cluster dimension grows for enlarged \eR, until the value
\eR=12 \aa \, for rhodopsin and \eR=9 \aa \, for AChE.
We notice, indeed, that for \eR = 12 \aa \, 
the degree distribution of rhodopsin exhibits two prominent maxima 
at k = 25 and 37, respectively (see Fig. 5), and so we have two 
different clusterizations. 
On the other hand, for the same value of \eR, AChE  
exhibits a degree distribution randomly spiked in the range 20 $<$ k 
$<$ 45.
For values of \eR  in the range $12 \div 25$ \aa, we have found a 
spreading of the distribution, which exhibits a series of spikes 
representing the fingerprint of the tertiary structure of the given
protein.
For values of \eR $>$ 25 \aa, the degree distribution is found to shrink
(see insert in Fig. 6) and, at  \eR = 80 \aa, all the nodes are 
found to be practically connected each other.
Here the degree distribution takes a delta-like shape centered 
at k=($\nu$-1). 
We conclude that,  \eR = 9 \aa, for AChE, and \eR = 12 \aa, for 
rhodopsin, should be taken as optimal values  to obtain the 
best resolution of the intimate protein structures. 
\par
From the above considerations, in the following 
we discuss four possible cases,
in which the interaction radius \eR, and the effective distance $D$ 
combine to produce different resolutions for the AA and AB directed model,
respectively. 
(Notice that the effective distance for the AB model has been assumed larger
than that for the AA model because of the finite size of the amino acid.)
\be
\begin{array}{lllll}
I.\hspace{1.2cm}   &D_{AA} \approx R, & D_{AB}\gtrsim R,
\hspace{.8cm}  & D^{\prime}_{AA} \gtrsim R, &     D^\prime_{AB}> R\\
II. \hspace{1.2cm}  &D_{AA} \lesssim  R, & D_{AB}\approx R,
\hspace{.8cm}  &       D^{\prime}_{AA}\gtrsim R, 
& D^\prime_{AB}\gtrsim R\\
III. \hspace{1.2cm}  &D_{AA} < R, & D_{AB}\lesssim R,
\hspace{.8cm}  &       D^{\prime}_{AA}\lesssim R, 
& D^{\prime}_{AB}\gtrsim R\\
IV. \hspace{1.2cm}  &D_{AA} \ll  R, & D_{AB} < R,
\hspace{.8cm}  &       D^{\prime}_{AA}\ll R, 
& D^{\prime}_{AB}< R
\end{array}
\ee
\par
Keeping in mind that the condition $D< R$ produces links and that, 
on the contrary, $D>R$ does not, the preceding cases are analyzed as follows:

Case I. Here the AA model discriminates different protein states  
better than the AB model.
\par
Case II. Here both the AA and the AB models are able to resolve well 
the two configurations.
In particular, the AA model is more sensitive to the change of the 
interaction radius.
\par
Case III. Here the AA model discriminates different protein states
worse than the AB model.
\par 
Case IV. Here it is rather difficult to discriminate the 
configurations both for the AA  and the AB models since the number of
links remains practically the same in both the configurations.
\par
The general trends discussed above can be quantitatively assessed 
for the proteins under test by selecting a significative set of  \eR
values for  rhodopsin and AChE, respectively.
Accordingly, for rhodopsin we take the \eR values
6, 12, 25, 50 \aa, and for AChE the \eR values 6, 9, 12, 25
\aa. 
The results of calculations are reported with an accuracy of three 
digits (see Appendix), which is considered appropriate for an 
experimental validation of the model. 

\par
Figures from 7 to 9 report the Nyquist plots of the global network
impedance, normalized to the value at zero frequency,  $Z(0)$, 
for the case of the engineered representations
of rhodopsin in the native state, Rho, and in the activated state,
Meta II. 
In all the figures the AA model is compared with the AB models
by adopting the same  convention for the symbols.
In each figure the plots corresponding to increasing values of  \eR 
are indicated as (a), (b), (c), (d) in clockwise orientation.
\par
As a general trend, the shape of the Nyquist spectra remains 
quite close to that of a semicircle, typical
of a single RC parallel impedance, except for small but significant
deviations from the semicircle when \eR = 6 \aa. 
\par 
Figure 7 shows the different impedance responses obtained with the AA 
model (tiny continuous line for Rho and dotted line for MetaII) and 
with the AB$_{\alpha, \alpha}$ directed model (bold continuous line for 
Rho, and dashed line for Meta II).
\par 
Figure 8 shows the Nyquist plots for the AA model and the AB$_{\alpha \beta,
\alpha \beta}$ directed model, respectively.
\par
Figure 9 shows the Nyquist plot for the AA model and the AB isotropic 
model. 
\par
For  \eR = 6 \aa, in all the cases the AA model exhibits a 
resolution between different configurations better than the AB models.
Thereby, the value \eR = 6 \aa \, pertains to case I: This value is a 
good choice but not the best.
\par
The value of  \eR = 12 \aa \,  provides the largest difference between 
the activated configuration and the native one.
Furthermore, we find that the AB$_{\alpha, \alpha}$ directed model 
(see Fig. 7), and the AB$_{\alpha \beta,
\alpha \beta}$ directed model (see Fig. 8) increase the  
differences, respectively of 7$\%$, and 5$\%$ with respect to the 
AA model. 
The value \eR = 12 \aa \, pertains to case II: The best value
for the AA model, a very good value for the AB directed model.
\par
For \eR = 25 \aa, the difference between the configurations starts to 
decrease for both the models, even if the directed AB model 
still exhibits a resolution increment with respect to that of
the AA model of 7$\%$, and 3$\%$ for its versions, AB$_{\alpha, \alpha}$
and AB$_{\alpha \beta, \alpha \beta}$, respectively. 
The value \eR = 25 \aa \, pertains to case III: 
For the AB models, this value of \eR is still a relevant one. 
\par
Finally, for \eR = 50 \aa,  the directed AB model exhibits a 
resolution increment with respect to that of the AA model, of 
3$\%$, and 1$\%$ for its versions, AB$_{\alpha, \alpha}$
and AB$_{\alpha \beta, \alpha \beta}$, respectively. 
The ability to resolve is decreasing but it remains significant for 
both the models. 
Accordingly the value of \eR =50 \aa \, is on the
boundary between case III and case IV.
\par
Figure 9 reports the comparison between the AA and the
isotropic AB model. 
We can observe, that for \eR $>$ 6 \aa the compared models give 
practically the same results, in other words, the AB isotropic model 
does not improve the AA model.
This outcome says that the AB isotropic model does not give to the 
C$_\beta$s an active role, unlike the AB directed model. 
With respect to the directed model it contains much more 
links, many of them slightly  varying in the conformational change.
This excess of (invariant) links hides the small differences 
between the native and active states. 
So, while the larger effects
due to C$_\alpha$ displacements can emerge once again, 
the smaller improvements
due to the C$_\beta$ displacements cannot be appreciated.
\par
By performing the same investigation for the case of AChE,
we noticed that the differences between the native and activated 
configurations are significantly smaller than those of rhodopsin.
In the AA model, with \eR = 6 \aa \, we have found a difference of 
only 6 $\%$, while, for rhodopsin it was of 16 $\%$.
Furthermore, this difference decreases at increasing \eR, unlike 
the case of rhodopsin. 
This implies a low  level of resolution between the configurations,
even for the most sensitive AB directed model. 
Accordingly, in Figs. 10 and 11  we report only the comparison 
between the AA model and the directed AB$_{\alpha, \beta}$ model, 
which among the AB models exhibits the best resolution.
\par 
From Fig. 10, we observe that for \eR = 6 \aa \, the AA model shows 
a resolution higher  than that of the AB model, as expected in case I. 
For \eR = 9 \aa \, the difference between the native and
activated configuration is less than 0.1 $\%$ in the AA model and it is of
2 $\%$ in the directed AB model, as expected in case III. 
Figure 10 (c) emphasizes this difference.
\par
Figure 11 reports the Nyquist plot of AChE for \eR = 12 \aa \, 
and 25 \aa, respectively. 
We notice that in both cases the AA model is no longer able to 
resolve the native from the activated configuration, 
while the directed AB model resolves a difference between the 
configurations of 2 $\%$ for \eR = 12 \aa, as expected in case III.
For \eR = 25 \aa, also the directed AB model is no longer able to 
resolve the difference between the configurations, as expected in 
case IV. 
\par
In Figs. 7 to 10, it is shown that for \eR = 6 \aa, the Nyquist 
plots take shapes, which are slightly squeezed and asymmetric
semicircles instead  of the perfect semi-circle pertaining to a single 
RC parallel circuit (see Appendix for details).
However, by increasing the value of \eR, the Nyquist plots better and better
approach the perfect semi-circle. 
The above peculiarities can be satisfactorily interpreted
in terms of  the Cole-Cole function \cite{Colecole} with one 
fitting parameter.
Accordingly, to interpret the Nyquist plot, we use 
the normalized dimensionless response function \cite{Macdonald}:
\be
I_{\omega}\,\equiv\, (Z(\omega)-Z_{\infty})/(Z(0)-Z_{\infty})
\ee  
which, in our case, reduces to $Z(\omega)/Z(0)$. 
The Cole-Cole fitting function is:
\be
I_{\omega}= \frac{1}{1+(i\omega\tau)^{1-\alpha}}, \hspace{.5cm} 
0\leq\alpha\leq1 \hspace{1cm} 
\ee
which leads to the following relation between the real and imaginary parts
of $I_{\omega}$
\be
({\cal R}(I_{\omega}) - \frac{a}{2})^2 + ({\cal I}(I_{\omega}) - \frac{b}{2})^2
= \frac{1}{c} (1-a) + \frac{1}{4}
\ee
with
$$
a= cos(\pi \alpha/2)
$$
$$
b= sin(\pi \alpha/2)
$$
$$
c=1 +(\omega \tau)^{2(1-\alpha)} + 2 (\omega \tau)^{1- \alpha}b
$$
where $1/ \tau=\omega_{M}$ is the frequency value corresponding to the 
maximum value taken by $-{\cal I}(I_{\omega})$ as function of 
${\cal R}(I_{\omega})$
(see Appendix).
\par
Figure 12 reports the fitting of the Nyquist plot for 2ACE, 
with \eR = 6 \aa \, in the AA model, obtained with the 
Cole-Cole function \cite{Colecole} and $\alpha=0.09$
together with the ideal semi-circle shape corresponding to $\alpha=0$.
The fact that for \eR $>$ 6 \aa \, the Nyquist plots take the ideal
semi-circle shape is explained by the predominance increase of parallel 
with respect to serial connections.
Thus, the network is no longer able to resolve the single relaxation times
pertaining to each RC link but exhibits an average time constant. 

The Cole-Cole function is one of the most used fitting functions in 
relaxation processes deviating from the Debye-Maxwell 
behavior \cite{Debye}. 
It is not the unique, of course, due the complexity of the possible 
origins of this deviation \cite{Davidsoncole,Macdonald}. 
On the other hand, it has been shown that its meaning is 
more vast than simple fitting function \cite{Metzler}. 
In fact, while the power spectrum $|I_{\omega}|^{2}$ associated with 
the Debye-Maxwell function is 
Lorentzian, and the correlation function is exponential, the power spectrum
associated with the Cole-Cole function is a more tangled object. 
It reduces 
to the Lorentzian distribution for $\alpha=0$, furthermore, it 
goes like $(\omega\tau)^{-2(1-\alpha)}$ for $\omega\tau\,\gg\,1$ and like
$(1+2(\omega\tau)^{(1-\alpha)}sin{(\pi\alpha/2))^{-1}}$ 
for $\omega\tau\,\ll\,1$.
The corresponding correlation function is the Mittag-Leffler function 
\cite{Erd} which interpolates between a stretched exponential pattern 
( $\omega\tau\,\gg\,1$) and an inverse power law decay  
($\omega\tau\,\ll\,1$). The exponent of both the functions is the 
same, 1-$\alpha$.

\par
For completeness,  in the Appendix we report the single resistance 
and capacitance values corresponding to the calculated Nyquist plots. 
\section{Conclusions}
We have carried out a systematic analysis of protein modelling 
by means of a coarse grained network approach.
As application we considered the case of bovine rhodopsin, the prototype
of the GPCR family, and the case of AChE enzyme, 
whose inihibition is one of the most qualified treatments of 
the Alzheimer disease.
\par
By using the topological features of the network, some relevant PDB 
entries  have been compared in terms of the number 
of links as function of the interacting radius.
We notice that the network map of the protein is able to distinguish among
different PDB entries, and also to reproduce, with a fine choice of 
the cut-off \eR, some topological properties of the protein structures.
We conclude that the network approach is a suitable tool to discriminate
among different protein structures. 
\par
By using the features of the impedance network associated with
the topological one, we have investigated the dynamic 
electrical response of the proteins through the Nyquist plot 
representation.
Accordingly, we predict for the rhodopsin an 
electrical response which is of a quite 
detectable level (up to difference of 22 \%
when passing from the native to the activated state).
Furthermore, a significative conformational change is identified both
with the one-node AA model and with the two-nodes AB models, 
the latter foreseing significantly larger 
differences among the configurations. 
These results are supported by some
experimental evidences \cite{Hou}: Bovine rhodopsin and the rat olfactory
receptor I7 has been immobilized
on a gold electrode building up a self-assembled multilayer. 
The electrochemical characteristic of these structures has been performed 
in a standard electrochemical cell, and has shown Nyquist plots qualitatively 
similar to those obtained in the present paper.
\par
Finally, we have found that the electrical responses of bovine rhodopsin
and torpedo AChE are quite different, with regard to
the possibility to distinguish
between the activated and native state. 
For the couple 2ACE-1VOT-2, the
maximal difference is of 6 \%, and it is obtained for a small 
value of the interaction radius \eR = 6 \aa. 
We conclude that such a small sensitivity is due to an effectively 
small difference between the two states. 
Furthermore, since the 
maximal resolution between the native and the activated state is for 
small values of \eR, the conformational change acts only among nearest 
neighboring.
However, this is not the only possibility. 
In fact, in a recent work on a particular enzyme, the cyclophilin
A (CypA) \cite{EIS05}, 
it has been shown that also in the native form, the 
enzyme lives part in the activated and part in the native state. 
In the activated state,
the percentage of activated configurations simply increases. 
According to this conclusion,
also for AChE, our results should demonstrate that the crystallographic 
image of native AChE is a mix of the native state and of the 
Huperzine A (in our case) state, and therefore its conformational is 
of weak relevance.       
\section{Acknowledgments}
The authors acknowledge the MIUR PRIN "Strumentazione elettronica integrata
per lo studio di variazioni conformazionali di proteine 
tramite misure elettriche" prot.2005091492.
We thank also dr.V. Akimov for useful discussions on the subject and 
prof. M. Barteri for his interesting comments.
\appendix*
\section{}
The frequency response of the impedance network presented here
can be modeled to a good degree of approximation, by the impedance of a 
single RC parallel circuit
\be
Z_{RC}= \frac{R}{1+i\omega RC} \equiv {\cal R}+i {\cal I}, 
\ee
where ${\cal R}, {\cal I}$ indicates the real and imaginary part of 
$Z_{RC}$, respectively.
It is convenient to normalize the impedance, in order to better 
compare results coming from different proteins and boundary 
conditions, so we introduce
\be
{\hat{Z}}_{RC}= 
\frac{Z_{RC}}{R} \equiv \hat{\cal{R}}+ i {\hat{\cal{I}}}, 
\label{A2}
\ee
with $R$ the impedance at zero frequency. 
Notice that the function (\ref{A2}) is a first order function, 
i.e. it has only one pole in frequency and it is stable. 
The maximum of $-\hat{\cal{I}}$ occurs for $-\hat{\cal{I}}= 
\hat{\cal{R}} = 1/2$ and is obtained for $\omega = \omega_{M} = 1/RC$.
\par
In the Nyquist plots calculated for the impedance networks, 
to account for the non-ideal semicircle shape, 
we have determined the frequency  $\omega^*$ at which $-\hat{\cal{I}}= 
\hat{\cal{R}}$ and defined an effective capacitance, $C^*$, by the relation
$\omega^* =1/RC^*$.   
The difference between the values of $C$ and
$C^*$ is a signature of the deviation of the Nyquist plot from the 
perfect semi-circle shape. 
The values of $R$, $C$, and $C^*$
corresponding to the Nyquist plots reported in Figs. 7, 8, 10, 11 are 
reported  
in Table I (rhodopsin) and in Table II (AChE). 
\vskip 1truecm


%
\newpage
\noindent
\begin{table}
\begin{center}
\footnotesize{
\caption{Resistance and capacitances of the RC 
single-impedance circuit equivalent to the protein impedance network.
Model AA is compared with the AB$_{\alpha,\beta}$ and
AB$_{\alpha\beta,\alpha\beta}$ models for \eR=6,12,25,50 \aa.
Rhodopsin is the analyzed protein }
\begin{tabular}{|c|c|c|c|c|c|c|c|c|c|} \hline
& \multicolumn{3}{|c|}{\bf AA Model} &\multicolumn{3}{c}{\bf 
AB$_{\alpha,\alpha}$ Model}
&\multicolumn{3}{|c|}{\bf AB$_{\alpha \beta, \alpha \beta}$ Model}\\ \hline\hline
\eR/Type &R ($P\Omega)$ &C(fF) &C$^*$(fF) &R($P\Omega$) & C(fF) &C$^*$ 
(fF) &R($P\Omega$) &C(fF) &C$^*$(fF) \\ \hline
& & & & & & & & &\\ 
\eR=6\aa & & & & & & & & &  \\   
Rho & 9.49 10$^{3}$ & 1.53 10$^{-4}$ & 1.32 10$^{-4}$ & 4.72 
10$^3$& 3.26 10$^{-4}$&
2.82 10$^{-4}$& 4.47 10$^3$ & 3.60 10$^{-4}$ & 3.19 10$^{-4}$ \\ 
Meta & 1.13 10$^{4}$ & 1.21 10$^{-4}$ & 9.31 10$^{-5}$ & 5.37 10$^3$
& 2.66 10$^{-4}$&
1.86 10$^{-4}$& 5.01 10$^3$ & 3.07 10$^{-4}$ & 2.50 10$^{-4}$ \\ \hline\hline
& & & & & & & & &\\ 
\eR=12\aa & & & & & & & & &  \\   
Rho & 2.51 10$^{2}$ & 5.37 10$^{-3}$ & 5.16 10$^{-3}$ & 1.29 10$^2$& 
9.69 10$^{-3}$&
9.69 10$^{-3}$& 9.07 10$^1$ & 1.47 10$^{-2}$ & 1.43 10$^{-2}$ \\ 
Meta & 3.22 10$^{2}$ & 3.57 10$^{-3}$ & 3.41 10$^{-3}$ & 1.81 10$^2$& 
5.51 10$^{-3}$&
5.01 10$^{-3}$& 1.24 10$^2$ & 9.35 10$^{-3}$ & 8.93 10$^{-3}$ \\ \hline\hline
& & & & & & & & &\\ 
\eR=25\aa & & & & & & & & &  \\   
Rho & 1.67 10$^{1}$ & 6.90 10$^{-2}$ & 6.90 10$^{-2}$ & 1.07 10$^6$& 
1.04 10$^{-1}$&
9.82 10$^{-2}$& 5.99  & 1.99 10$^{-1}$ & 1.96 10$^{-1}$ \\ 
Meta & 2.04 10$^{1}$ & 5.38 10$^{-2}$ & 5.38 10$^{-2}$ & 1.42 10$^6$& 
7.04 10$^{-2}$&
7.04 10$^{-2}$& 7.06  & 2.83 10$^{-1}$ & 2.83 10$^{-1}$ \\ \hline\hline
& & & & & & & & &\\ 
\eR=50\aa & & & & & & & & &  \\   
Rho & 2.31  & 4.32 10$^{-1}$ & 4.32 10$^{-1}$ & 1.61 & 5.65 10$^{-1}$&
5.65 10$^{-1}$& 0.82  & 1.21  & 1.21  \\ 
Meta & 2.72  & 3.34 10$^{-1}$ & 3.34 10$^{-1}$ & 1.97 & 4.24 10$^{-1}$&
4.24 10$^{-1}$& 0.98  & 9.30 10$^{-1}$ & 9.30 10$^{-1}$ \\ \hline\hline
\end{tabular}
}
\end{center}
\end{table}
\begin{table}
\begin{center}
\footnotesize{
\caption{Resistance and capacitances of the RC 
single-impedance circuit equivalent to the protein impedance network. 
Model AA is compared with the AB$_{\alpha,\alpha}$ and
AB$_{\alpha\beta,\alpha\beta}$ models, for \eR=6,9,12,25 \aa.
AChE is the analyzed protein }
\begin{tabular}{|c|c|c|c|c|c|c|} \hline
& \multicolumn{3}{|c|}{\bf AA Model} &\multicolumn{3}{|c|}{\bf 
AB$_ {\alpha , \beta}$
Model}\\ \hline\hline
\eR/Type &R($P\Omega)$ &C(fF) &C$^*$(fF) &R($P\Omega$) & C(fF) &
C$^*$ (fF)  \\ \hline
& & & & & &  \\ 
\eR=6\aa & & &  & & &   \\ 
2ACE & 1.51 10$^{4}$ & 7.23 10$^{-5}$ & 6.00 10$^{-5}$ 
& 3.63 10$^{3}$ & 3.93 10$^{-4}$& 3.67 10$^{-4}$  \\
1VOT-2 & 1.43 10$^{4}$ & 8.74 10$^{-5}$ & 6.99 10$^{-5}$ & 
3.50 10$^{3}$ & 4.08 10$^{-4}$& 3.57 10$^{-4}$  \\ \hline \hline
& & & & & &  \\ 
\eR=9\aa & & &  & & &   \\ 
2ACE & 1.45 10$^{3}$ & 1.11 10$^{-3}$ & 1.01 10$^{-3}$ & 5.76 
10$^{2}$ & 2.89 10$^{-3}$& 2.67 10$^{-3}$  \\
1VOT-2 & 1.46 10$^{3}$ & 1.10 10$^{-3}$ & 1.07 10$^{-3}$ & 5.88 
10$^{2}$ & 2.84 10$^{-3}$& 2.62 10$^{-3}$  \\ \hline \hline
& & & & & &  \\ 
\eR=12\aa & & &  & & &   \\ 
2ACE & 3.57 10$^{2}$ & 5.10 10$^{-3}$ & 5.10 10$^{-3}$ & 1.42 
10$^{2}$ & 1.28 10$^{-2}$& 1.28 10$^{-2}$  \\
1VOT-2 &  3.57 10$^{2}$ & 5.10 10$^{-3}$ & 5.10 10$^{-3}$ & 1.45 
10$^{2}$ & 2.51 10$^{-2}$& 2.51 10$^{-2}$ \\ \hline \hline
& & & & & &  \\ 
\eR=25\aa & & &  & & &   \\ 
2ACE & 2.08 10$^{1}$ & 9.80 10$^{-2}$ & 9.80 10$^{-2}$ & 9.71  & 2.07 
10$^{-1}$& 2.06 10$^{-1}$  \\
1VOT-2 &  2.08 10$^{1}$ & 9.63 10$^{-2}$ & 9.63 10$^{-2}$ & 9.76  
& 2.05 10$^{-1}$& 2.05 10$^{-1}$  \\ \hline \hline
\end{tabular}
}
\end{center}
\end{table}

\begin{figure}
\includegraphics{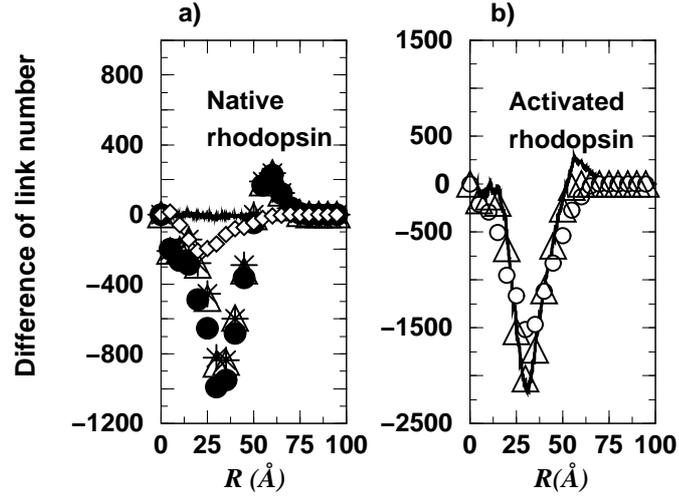}
\caption{Link number difference (LND) for different  configurations
of rhodopsin versus the interaction radius. 
All the configurations contain the same  number of amino acids.
Figure 1 (a): Native rhodopsin. 
Stars refer to the LND between the a-chain
of 1U19, 1U19$a$ and the engineered representation of native rhodopsin, 
Rho. Open triangles refer to the LND between the a-chain of 
2G87, 2G87$a$ and Rho. 
Full circles refer  to the LND between the a-chain of 2HPY, 2HPY$a$ and Rho.
Open diamonds refer to the LND between the a-chains 2HPY$a$ and 1U19$a$.
Continuous line refers to the LND between 2G87${a}$ and  1U19$a$.
Figure 1(b): Activated rhodopsin. 
Continuous line refers to the LND between the engineered representation 
of rhodopsin in light, MetaII, and  Rho. 
Open circles refer to the LND between 2I37  and Rho. 
Open triangles refer to the difference between 1LN6  and Rho. 
In the second and third case the Rho sequence has been deprived of the 
amino acids that are not present in 2I37 and 1LN6, respectively.}
 \end{figure}
\newpage

\begin{figure}
\includegraphics{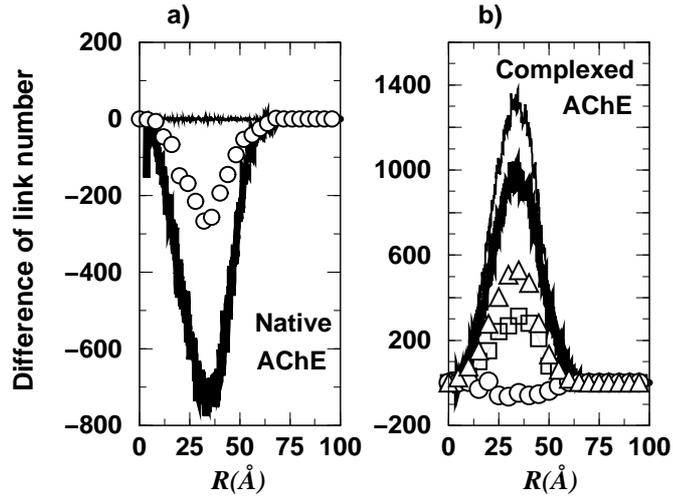} 
\caption{
Link number difference (LND) for different  configurations of AChE 
versus the interaction radius. 
All the configurations contain the same  number amino acids.
Figure 2(a): Native configurations.  
Tiny continuous line refers to the
LND between two chains of the 1EA5 representation: 1EA5$b$, 1EA5$a$.
Bold continuous line refers to the difference  between 1W75$a$ and 
2ACE. 
Open circles refer to  both: the LND between 1EA5$a$ and 1W75$b$ 
and  the LND between 1EA5$b$ and 1W75$b$.
Figure 2(b): Complexed configurations.
Tiny black line refers to the LND between 2ACE and 1ACJ. 
Bold black line refers to the LND between 2ACE and 1ACL.
Open squares  refer to the LND between 2ACE and 1AX9. 
Open circles refer to the LND between 2ACE and 1VOT.
Open triangles refer to the LND between 2ACE and 1GPK$a$. }
\end{figure}
\newpage

\begin{figure}
\includegraphics{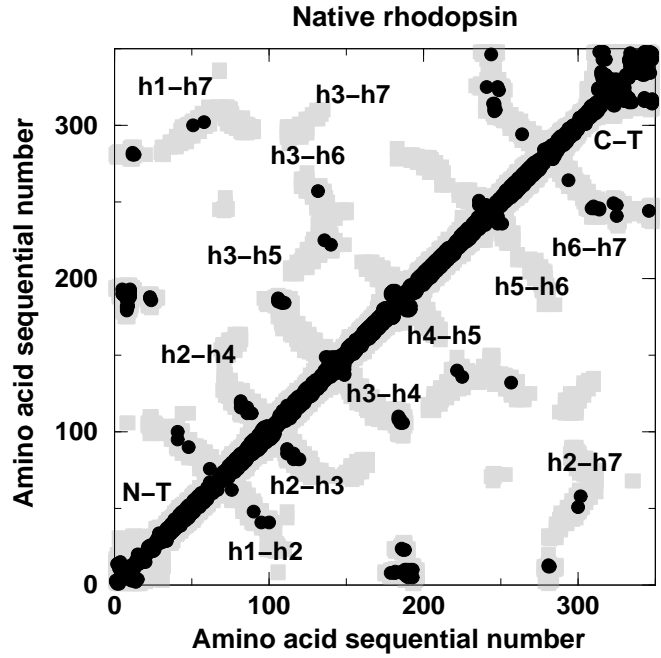}
\caption{
Adjacency matrix of native rhodopsin, Rho. 
The x,y axes report the sequential number of the amino acids. 
Black full circles refer to \eR = 6 \aa, and grey open squares 
to \eR = 12 \aa. 
Each circle/square correspond to a link between the couple
of amino acids (x,y).
The main domains associated with
the connections among closest helices are explicitly indicated.}
\end{figure}
\newpage

\begin{figure}
\includegraphics{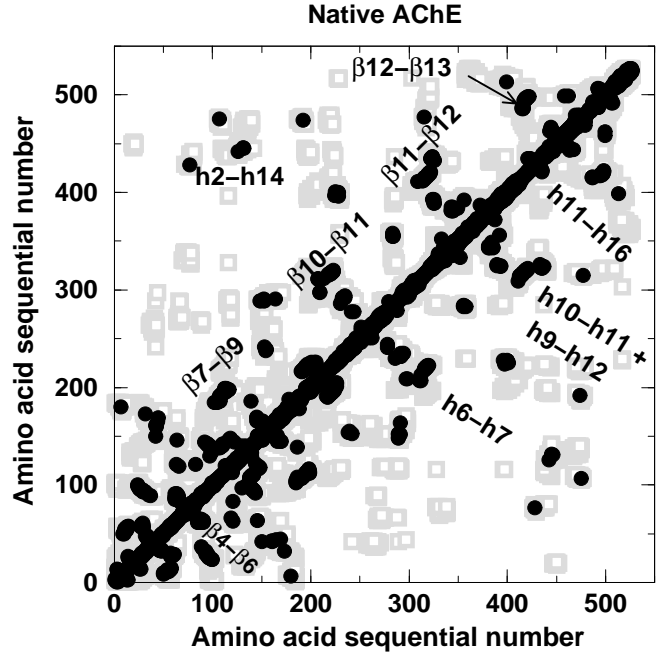}
\caption{
Adjacency matrix of 2ACE. 
The x,y axes report the sequential number of the amino acid. 
Black full circles refer to \eR = 6 \aa, and 
grey open squares to \eR = 12 \aa. 
Each circle/square corresponds to a link between the couple
of amino acids (x,y). 
Dashed lines are in correspondence with the three
amino acids of the active site.} 
\end{figure}
\newpage
\begin{figure}
\includegraphics{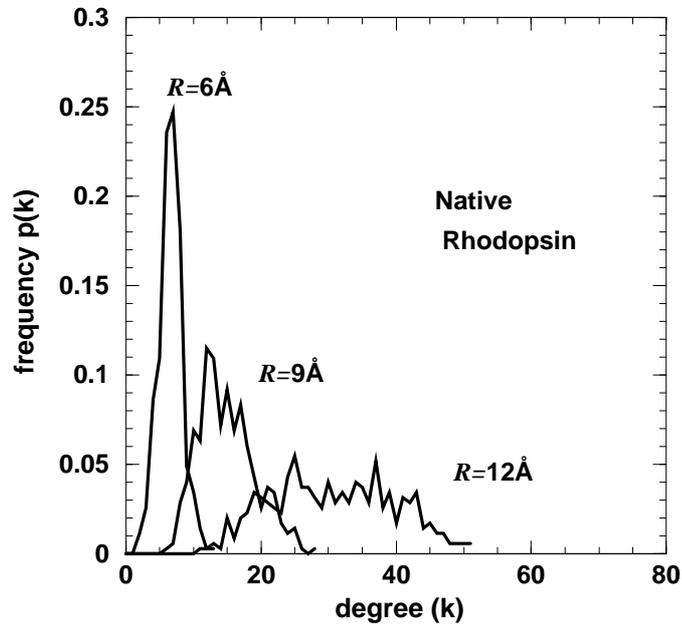}
\caption{ Degree distribution of native-rhodopsin (rho) 
network for increasing values of the interaction radius \eR. 
}
\end{figure}
\begin{figure}
\includegraphics{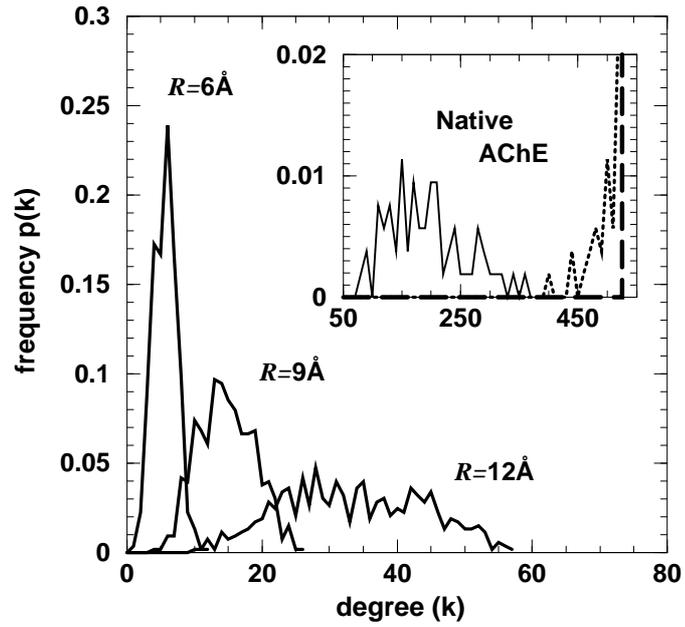}
\caption{ Degree distribution of the native-AChE network for increasing 
values of the interaction radius \eR. 
The inset reports the degree distribution
for \eR=25 \aa (solid line), \eR=50 \aa (dotted line),
\eR =80 \aa (bold dashed line).}
\end{figure}
\newpage

\begin{figure}
\includegraphics{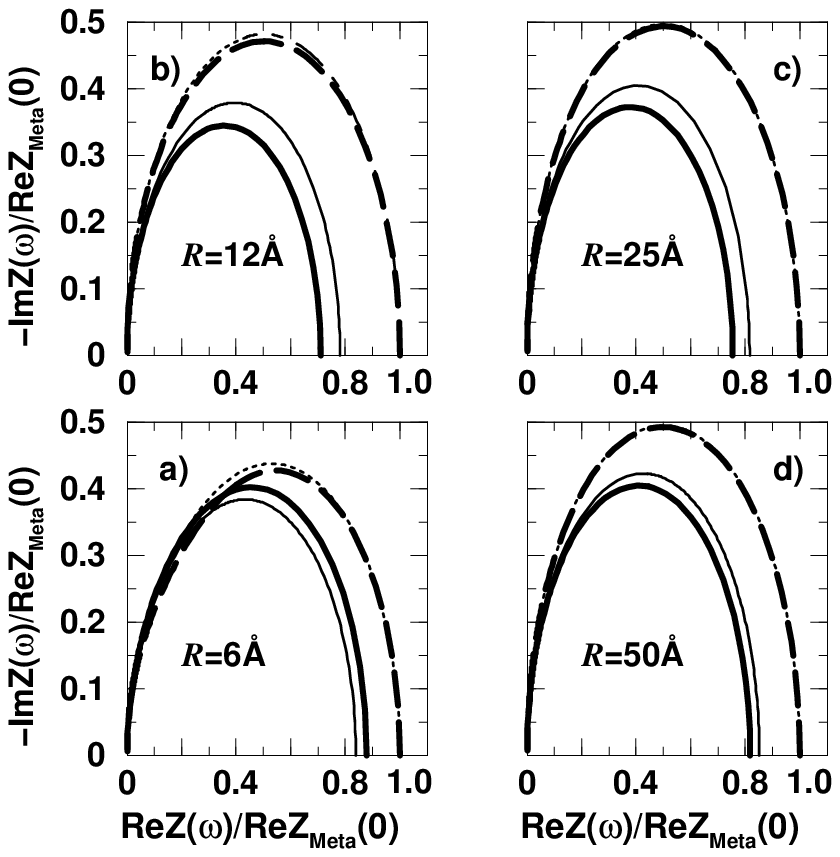}
\caption{
Nyquist plots of the network impedance associated with the protein. 
Native rhodopsin  is compared with activated rhodopsin.
The protein sequences are  the engineered Rho (native) and Meta II 
(activated). 
For the AA model:
Continuous tiny lines refer to Rho, dotted lines refer
to Meta II. 
For the AB$_{\alpha,\alpha}$ directed model: 
Bold continuous lines refer to Rho, 
dashed lines refer to Meta II.  
Plots are reported for increasing  values of the interaction radius
in the range  from 6 \aa to 50 \aa following a clock-wise orientation.
}
\end{figure}
\newpage

\begin{figure}
\includegraphics{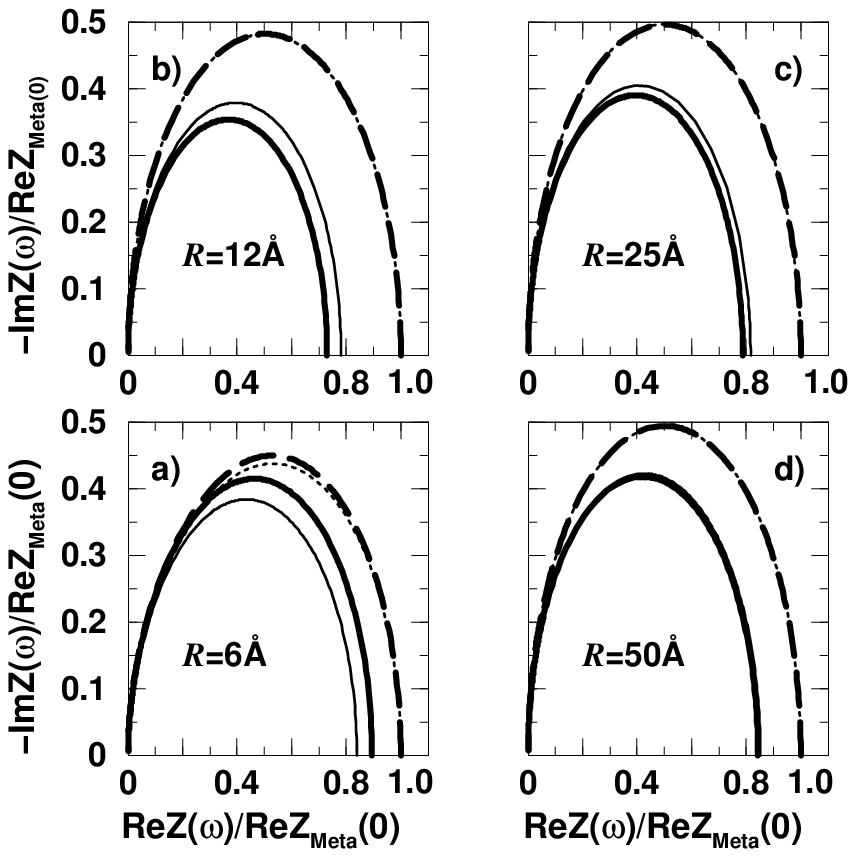}
\caption{
Nyquist plot of the network impedance associated with the protein. 
Native rhodopsin is compared with activated rhodopsin.
The protein sequences are  the engineered Rho (native) and Meta II 
(activated).
For the AA Model:
Continuous tiny lines refer to Rho, and dotted lines refer
to Meta II. 
For the AB$_{\alpha\beta,\alpha\beta}$ directed model:
Bold continuous lines refer to Rho, and dashed lines refer to Meta II.
Plots are reported for increasing  values of the interaction radius
in the range  from 6 \aa \, to 50 \aa \, following a clock-wise orientation.
}
\end{figure}
\newpage

\begin{figure}
\includegraphics{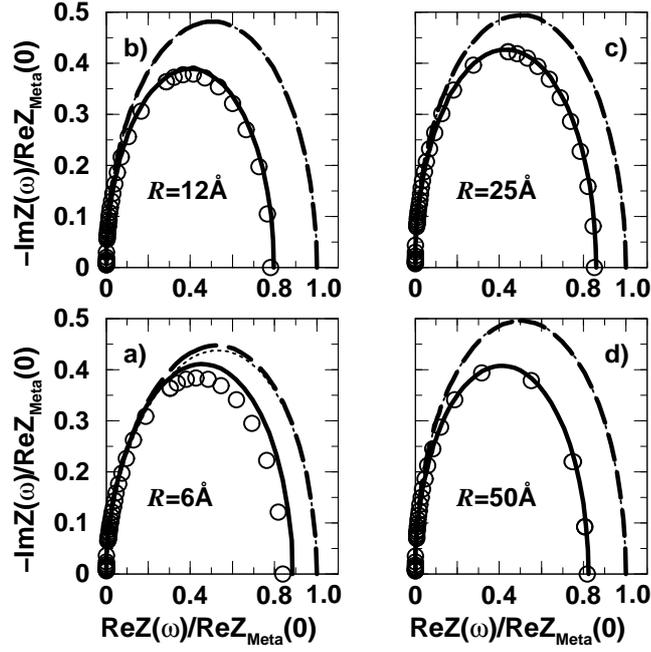}
\caption{
Nyquist plot of the network impedance associated with the protein. 
Native rhodopsin is compared with activated rhodopsin.
The protein sequences are  the engineered Rho (native) and Meta II 
(activated).
For the AA Model: Circles refer to Rho, and dotted lines refer
to Meta II. 
For the AB isotropic model (with in-contact on the first C$_\alpha$ 
carbon atom and out-contact on the last C$_\alpha$ carbon atom): 
Bold continuous lines refer to Rho, and dashed lines refer to Meta II. 
Plots are reported for increasing  values of the interaction radius
in the range  from 6 \aa \, to 50 \aa \, following a clock-wise orientation.
}
\end{figure}
\newpage

\begin{figure}
\includegraphics{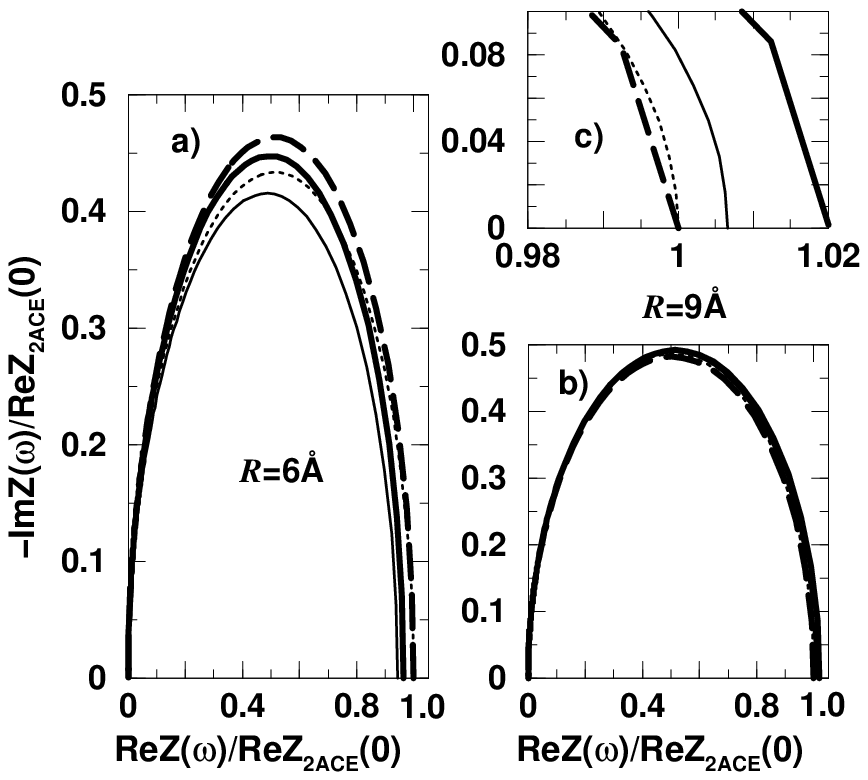}
\caption{
Nyquist plot of the network impedance associated with the protein.  
Native AChE, 2ACE, is compared with AChE complexed with 
Huperzyne A, 1VOT-2. 
For the AA model:
Tiny continuous lines refer to 1VOT-2, dotted lines refer to 2ACE. 
For the AB$_{\alpha,\beta}$ directed model: Bold continuous lines 
refer to 1VOT-2, 
and dashed lines refer to 2ACE in the directed AB$_{\alpha,\beta}$ model. 
In Fig. 9(a)  \eR = 6 \aa, while in Figs. 9(b) and 9(c) \eR = 9 \aa;
Fig. 9(c) is a zoom of Fig. 9(b).
}
\end{figure}
\newpage

\begin{figure}
\includegraphics{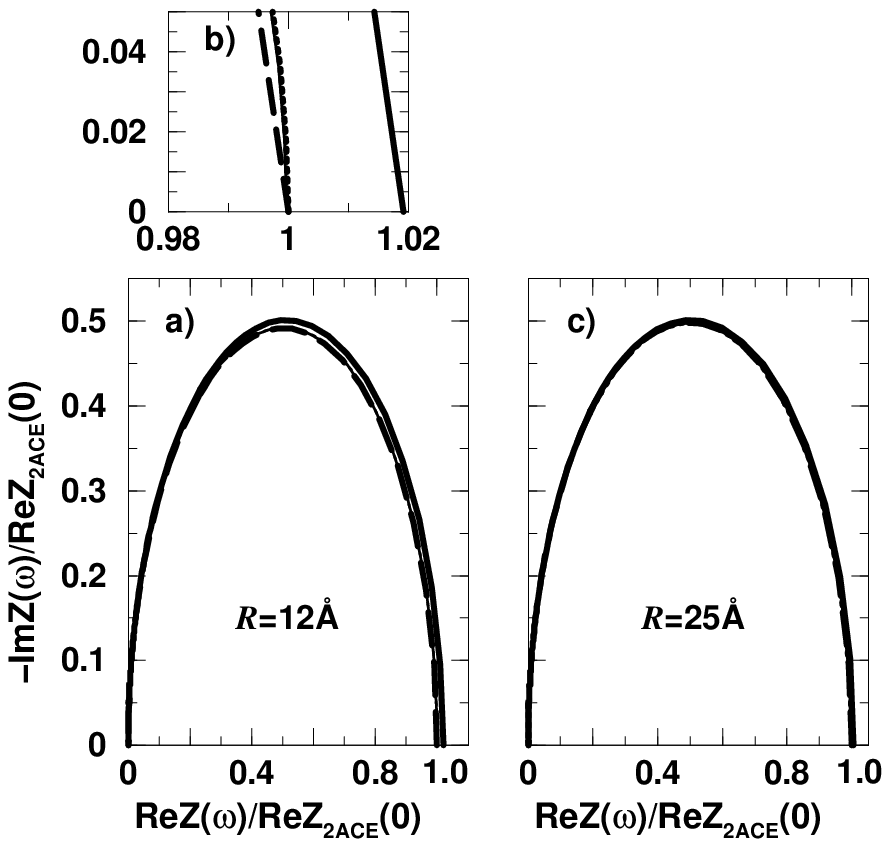}
\caption{
Nyquist plot of the  network impedance associated with the protein. 
Native AChE, 2ACE, is compared with AChE complexed with 
Huperzyne A, 1VOT-2. 
For the AA model:
Tiny continuous lines refer to 1VOT-2, dotted lines refer to 2ACE. 
For the AB$_{\alpha,\beta}$ directed model: Bold continuous lines refer to 1VOT-2, 
and dashed lines refer to 2ACE in the directed AB$_{\alpha,\beta}$ model.
In Figs. 10(a) and 10(b)  \eR = 12 \aa, and Fig. 10(b) is a zoom of 
Fig. 10(a). In Fig. 10(c) \eR = 25 \aa. }
\end{figure}
\begin{figure}
\newpage
\includegraphics{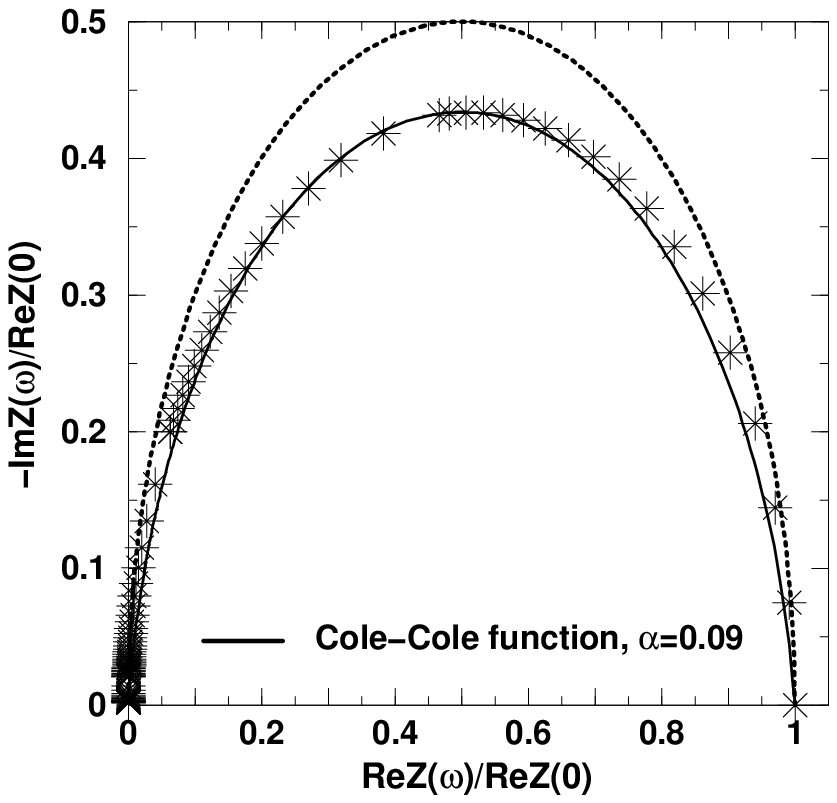}
\caption{
Nyquist plot of the 2ACE for \eR = 6 \aa, 
with the Cole-Cole distribution. 
Stars refer to data from simulations, the dotted
line refers to the Debye distribution function, 
the continuous line refers to the Cole-Cole distribution function 
with the best fit parameter $\alpha=0.09$. 
For all the distributions the value $\tau$ =1.12 s is used.
}
\end{figure}

\begin{thebibliography}{99}
\bibitem{Dutton}
C. C. Moser, J. M. Keske, K. Warncke, R. S. Farid, and P. L. Dutton, 
\Journal{Nature}{355}{796}{1992} and references therein.
%
\bibitem{Weiss}
T. Fisher Weiss, {\em Cellular Biophysics}, Vol. 2, 2th ed. (The MIT Press, Cambridge, Massachusetts, 1997).
%
\bibitem{Frau}
H. Fraufelder, P. G. Wolynes, and R. H. Austin, \Journal{\RMP}{71}{S419}{1999}.
%
\bibitem{Onuchic}
J. N. Onuchic, D. N. Beratan, J. R. Winkler, and H. B. Gray, 
\Journal{Annu. Rev. Biophys. Biomol. Struct.}{21}{349}{1992};
V. S. Pande, and J. N. Onuchic, \Journal{\PRL}{78}{146}{1997}.
%
\bibitem{SHE00}
S.-Y. Sheu, D.-Y. Yang, H. L. Selzle, and E. W. Schlag, \Journal{\EPJD}{20}{557}{2002};
E. W. Schlag, S.-Y. Sheu, D.-Y. Yang, H. L. Selzle, and S. H. Lin, \Journal{
J. Phys. Chem.}{104}{7790}{2000}.
%
\bibitem{rinaldi03}
R. Rinaldi, A. Biasco, G. Maruccio, R. Cingolani, D. Alliata, 
L. Andolfi, P. Facci, F. De Rienzo, R. Di Felice, E. Molinari, 
M. Verbeet, and G. Canters,  \Journal{Appl. Phys. Lett.}{82}{472}{2003}. 
%
\bibitem{jin06}
Y. Jin, N. Friedman, M. Sheves, T. He, and D. Cahen, 
\Journal{PNAS}{103}{8601}{2006}.
%
\bibitem{Din01}
T. V. Dinh, B. M. Cullum, and D. L. Stokes, \Journal{\SAB}{74}{2}{2001}.
%
\bibitem{PDB}
H. M. Berman, J. Westbrook, Z. Feng, G. Gilliland, T. N. Bhat, 
H. Weissing, I. N. Shindyalov, and P. E. Bourne, \Journal{Nucleic Acids Research}{28}{235}{2000}.
%
\bibitem{Go}
A. McCammon, and S. C. Harvey, {\em Dynamics of Proteins and Nucleic 
Acids} (Cambridge University Press, 1987); A. Kitao, and N. Go, 
\Journal{Curr. Opin. Struct. Bio.}{9}{164}{1999};  P. Carloni, 
U. Rothlisberger, and M. Parrinello, \Journal{Acc. Chem. Res.}{35}{455}{2002}.
%
\bibitem{Ben}
D. ben-Avraham, 
\Journal{\PRB}{47}{14559}{1993}; C. Micheletti, G. Lattanzi, and 
A. Maritan, 
\Journal{J. Mol. Bio.}{321}{909}{2002}. 
%
\bibitem{Tirion}
M. Tirion, \Journal{\PRL}{77}{1905}{1996}.
%
\bibitem{Atilgan}
A. R. Atilgan, S. R. Durell, R. L. Jernigan, M. C. Demirel, O. Keskin, and I. Bahar,
\Journal{\BiJ}{80}{505}{2001}; 
M. K. Kim, R. L. Jernigan, and G. S. Chirikjian, \Journal{\BiJ}{89}{43}{2005}.
%
\bibitem{Maritan}
C. Micheletti, G. Lattanzi, and A. Maritan, \Journal{\JMB}{321}{909}{2002};
C. Micheletti, P. Carloni, and A. Maritan, \Journal{Proteins}{55}{635}
{2004}.
%
\bibitem{UPON04}
E. Alfinito, V. Akimov, C. Pennetta, and L. Reggiani, in {\em Unsolved 
Problems of Noise and Fluctuations,} {\em AIP Conference Proceedings 
800}, edited by L. Reggiani, C. Pennetta, V. Akimov,  E. Alfinito, 
and M. Rosini (AIP, Melville, New York, 2005), pp.381-387.
%
\bibitem{WYL05}
C. Pennetta, V. Akimov, E. Alfinito, L. Reggiani, T. Gorojankina, 
J. Minic, E. Pajot-Augy, M. A. Persuy, et al, in {\em Nanotechnologies of the Life Science,4} edited by Challa S. S. R. Kumar
(Wiley-VCH, Weinheim, 2006), pp. 217-240.
%
\bibitem{ICNF05}
C. Pennetta, V. Akimov, E. Alfinito, L. Reggiani, and G. Gomila,
in 
 {\em Noise and Information in Nanoelectronics, Sensors and Standards 
II- Proceedings of SPIE 5472}, edited by J. M. Smulko, Y. Blanter, M. I. Dykman, and L. B. Kish (Int. Soc. Opt. Eng. Bellingham, 2004) 
pp. 172-182. 
%
\bibitem{Yang}
H. Yang, G. Luo, P. Karnchanaphanurach, T. M. Louie, I. Rech, S. Cova, 
L. Xun, and X. S. Xie, \Journal{\em Sciences}{302}{262}{2003}.
%
\bibitem{Song}
X. Song, \Journal{\JCP}{116}{9359}{2002}.
%
\bibitem{Rammal}
R. Rammal, C. Tannous, and A. M. S. Tremblay, \Journal{Phys. Rev. A}{31}{2662}{1985}.
%
\bibitem{Odor}
G. \`Odor, \Journal{Rev. Mod. Phys.}{76}{663}{2004} and references therein.
%
\bibitem{Gether}
U. Gether, and B. K. Kobilka,
\Journal{J. Bio. Chem.}{273}{17979}{1998};
R. J. Lefkowitz, \Journal{ Nature Cell Biology}{2}{E133}{2000};
K. Palczewski, T. Hori, C. A. Behnke, H. Motoshima, B. A. Fox, 
I. Le Trong, D. C. Teller, T. Okada, R. E. Stenkamp, M. Yamamoto, and 
M. Miyano, \Journal{Science}{289}{739}{2000}.
%
\bibitem{Santosh}
S. T. Menon, M. Han, and T. P. Sakmar, \Journal{Physio. Rev.}{81}{1659}{2001}.
%
\bibitem{Barabasi}
R. Albert, and A. L. Barabasi,
\Journal{Rev. Mod. Phys.}{74}{47}{2002}.
%
\bibitem{Hou}
Y. Hou, N. Jaffrezic-Renault, C. Martelet, A. Zhang, J. Minic-Vidic, T. Gorojankina, M-A. Persuy, E. Pajot-Augy, et al., \Journal{\Bios}{21}{1393}{2006};
Y. Hou, S. Helali, A. Zhang, N. Jaffrezic-Renault, C. Martelet, 
J. Minic, T. Gorojankina, M-A. Persuy, et al.\Journal{\Bios}{22}{1550}{2007}.
%

%
\bibitem{Colecole}
K. S. Cole, and R. H. Cole, \Journal{\JCP}{9}{341}{1941}.
%
\bibitem{Macdonald}
J. R. Macdonald,\Journal{\JAP}{62}{R51}{1987} and references therein. 
%
\bibitem{Debye}
P. Debye, {\em Polar Molecules} (The Chemical Catalogue Company, 
New York, 1929).
%
\bibitem{Davidsoncole}
D. W. Davidson, and R. H. Cole, \Journal{\JCP}{19}{1417}{1951};
A. K. Jonscher, \Journal{Nature}{267}{673}{1977};
K. L. Ngai, \Journal{\PRB}{22}{2066}{1980};
V. Raicu, \Journal{\PRB}{60}{4677}{1999}.
%
\bibitem{Metzler}
R. Metzler, and J. Klafter, \Journal{J. Non-Crystal. Sol.}{305}{81}{2002}; 
K. Weron, and M. Kotulski, \Journal{Physica A}{232}{180}{1996}.
%
\bibitem{Erd}A. Erd\'elyi (Ed.), {\em Tables of Integral 
Transformations} Bateman Manuscript Project, Vol.1 (McGraw-Hill, New York, 
1954)
%
\bibitem{EIS05}
E. Z. Eisenmesser, O. Millet, W. Labeikovsky, D. M. Korzhnev,
M. Wolf-Wartz, D. A. Bosco, J. J. Skalicky, L. E. Kay, and D. Kern, 
\Journal{\NLE}{483}{117}{2005}.
%
%

%
\end{thebibliography}
\end{document}